\documentclass[conference]{IEEEtran}

\makeatletter

\newcommand{\Rmnum}[1]{\expandafter\@slowromancap\romannumeral #1@}
\makeatother
\usepackage{url}
\usepackage{amsmath,bm}
\usepackage{amssymb}
\usepackage{extarrows}
\usepackage{algorithm}
\usepackage{graphicx} 
\usepackage{epstopdf}
\usepackage{amsthm}
\usepackage{multirow}
\usepackage{algpseudocode}
\usepackage{multicol}
\usepackage{color}
 \usepackage{cite}
\usepackage{setspace} 
\usepackage{caption}
\usepackage{subcaption}
\usepackage{subfig}
\usepackage{makecell}
\usepackage{stfloats}
\newcommand\blfootnote[1]{%
  \begingroup
  \renewcommand\thefootnote{}\footnote{#1}%
  \addtocounter{footnote}{-1}%
  \endgroup
}
\makeatletter

\renewcommand{\algorithmicrequire}{ \textbf{Input:}} 
\renewcommand{\algorithmicensure}{ \textbf{Output:}} 
\theoremstyle{plain}

\algdef{SE}[DOWHILE]{Do}{doWhile}{\algorithmicdo}[1]{\algorithmicwhile\ #1}%
\begin{document}
\include{com.tex}

\title{Environment-Aware Diffusion Model for Massive MIMO-OFDM Channel Estimation}
\author{\IEEEauthorblockN{Wanchen Hu$^{*}$, Jie Yang$^{*}$, Yi Song$^{\dagger}$, Jun Xia$^\ddagger$, Shuangyang Li$^{\dagger}$, Yu Zhu$^{*}$, and Giuseppe Caire$^{\dagger}$\\
	\IEEEauthorblockA{$^{*}$ Fudan University, Shanghai, China,  
    Emails: \{huwc23@m., yangjie23@m., zhuyu@\}fudan.edu.cn\\
        $^{\dagger}$ Technische Universitat, Berlin, Germany.
        Emails: \{yi.song, shuangyang.li, caire\}@tu-berlin.de \\
        $^{\ddagger}$ The Hong Kong University of Science and Technology (Guangzhou), Guangzhou, China.\\
        Email: junxia@hkust-gz.edu.cn\\
        (Invited paper)
    }
}
}

\maketitle
\blfootnote{
This work is partly supported by the National Natural Science Foundation of China under Grant No. 62471145, and the Natural Science Foundation of Shanghai under Grant No. 23ZR1407300. The work of Jun Xia is supported by the National Natural Science Foundation of China Project (No. 623B2086), CCF-GHFund (No. OF 2026005), CIPS-SMP-Zhipu Large Model Fund, Ant Group, and TeleAI of China Telecom. The work of Shuangyang Li was supported in part by European Research Council (ERC) under ERC Starting [Foundations of Delay Doppler Communications and Sensing (FUNDOCS)] under Grant 101220383. The work of Giuseppe Caire was supported by the Gottfried Wilhelm Leibniz-Preis 2021 of German Science Foundation (DFG).
}
\begin{abstract}
This paper proposes an environment-aware diffusion based channel estimation in massive multiple-input multiple-output orthogonal frequency division multiplexing (MIMO-OFDM) systems. The high dimensionality of massive MIMO channels combined with limited pilot resources makes accurate estimation challenging. To address this issue, we exploit the spatial variability of wireless channels by training a diffusion model to learn the location-conditioned distribution of channel state information, which provides an environment-aware prior for channel estimation. Based on this learned prior, a posterior inference algorithm is developed to incorporate pilot observations into the reverse diffusion process, enabling Bayesian channel estimation by combining the received-signal likelihood with the learned channel prior. By jointly leveraging location information and measurement data, the proposed approach improves estimation accuracy under limited pilot resources. Simulation results based on ray-tracing channel datasets demonstrate that the proposed method consistently outperforms conventional estimators and existing learning-based approaches across various signal-to-noise ratios and pilot configurations.


\end{abstract}

\begin{IEEEkeywords}
MIMO, environment-aware channel estimation, deep learning, diffusion model
\end{IEEEkeywords}

%
\IEEEpeerreviewmaketitle

\section{Introduction}

The advent of sixth-generation (6G) wireless networks imposes large demands on accurate channel state information (CSI), which is fundamental to enabling ultra-high data rates, massive connectivity, and ultra-reliable communications \cite{kin2023towards}. However, acquiring accurate CSI becomes increasingly challenging as wireless systems evolve toward large-scale antenna arrays and more complex propagation environments \cite{zeng2021towardEnvironment}. Specifically, the pilot overhead required for channel estimation grows rapidly with the number of antennas, making pilot resources increasingly insufficient in massive  multiple-input multiple-output (MIMO) systems. On the other hand, the complicated propagation conditions in realistic environments make it difficult to accurately characterize the statistical properties of wireless channels using conventional analytical models \cite{qiu2025AI}. In massive MIMO systems employing orthogonal frequency division multiplexing (OFDM), the wideband channel is observed across multiple subcarriers and exhibits structured correlations in both spatial and frequency domains \cite{hu2021phase}. These correlations originate from the same set of underlying multipath components with limited support in angle and delay. Since the multipath parameters are unknown and random in practical environments, the resulting channel distribution becomes a high-dimensional non-Gaussian random variable with complex statistics, making accurate prior modeling challenging.

In practical propagation environments, wireless channels are strongly coupled with the surrounding physical environment, including the geometry of buildings, scattering objects, and therefore vary with the user location. 
Studies have shown that incorporating prior environmental information can effectively improve system performance while reducing training overhead \cite{hu2025QNN}, \cite{kim2026large}. For example, the channel knowledge map (CKM)  maps the user location to channel knowledge and has been applied in many applications such as codebook design \cite{wei2026ckm} and resource allocation \cite{yue2024channel}.  While CKM provides a useful representation framework, effectively leveraging such location-dependent channel information for high-accuracy channel estimation remains challenging, particularly due to the complex distribution of wireless channels.

To address this challenge, powerful data-driven models capable of learning complex channel distributions are required. In recent years, generative models have emerged as effective tools for modeling high-dimensional data distributions \cite{yang2025llm}. Among them, diffusion models (DMs) have attracted significant attention due to their strong distribution modeling capability and stable training behavior compared with generative adversarial networks (GANs) and variational autoencoders (VAEs) \cite{ho2020denoising}. By progressively corrupting data with Gaussian noise and learning the corresponding reverse denoising process, diffusion models can effectively capture complex data distributions. Recently, diffusion models have been applied to wireless channel estimation as generative priors \cite{fesl2024DiffusionBasedGenerativePrior}, \cite{zhou2025Generative}, demonstrating promising potential. However, existing studies typically assume a global prior distribution that is independent of the environment, ignoring the spatially varying channel statistics.

In this paper, we propose an environment-aware diffusion model (EA-DM) for MIMO-OFDM channel estimation by exploiting the spatial variability of wireless channels. We first analyze the spatial variability of channel statistics under a fixed environment and demonstrate that channel features vary significantly across different user locations. To effectively capture such complex channel distributions, we develop a diffusion-based channel estimation framework built upon the Diffusion Transformer (DiT) architecture. The proposed model learns location-dependent channel priors and integrates them into the reverse diffusion process to guide channel reconstruction from limited pilot observations. Numerical results demonstrate that the proposed method achieves superior estimation accuracy compared with conventional approaches that rely on location-independent channel priors, highlighting the benefit of incorporating environment-aware channel priors into channel estimation.

\emph{Notations:} Boldface uppercase (lowercase) letters denote matrices, e.g., $\mathbf{A}$, (column vectors, e.g., $\mathbf{a}$), and italic letters denote scalars, e.g., $a$. $\mathbb {C}^{M\times N}$ ($\mathbb {C}^{M}$) and $\mathbb {R}^{M\times N}$ ($\mathbb {R}^{M}$) denote the sets of $M\times N$ ($M\times 1$) complex and real matrices (vectors), respectively.  
$(\cdot)^{\rm T}$ and $(\cdot)^{\rm H}$ stand for the transpose and conjugate transpose, respectively. $\mathbf{I}_N$ represents the $N \times N$ identity matrix. $\otimes$ denotes the Kronecker product. $\left\| \cdot \right\|_2$ denotes the 2-norm. $\mathcal{N}\left(\mathbf{x} ;\mathbf{a},\boldsymbol{\Sigma} \right)$ denotes the Gaussian density for $\mathbf{x}$ with mean $\mathbf{a}$ and covariance $\boldsymbol{\Sigma}$. 




\section{System Model and Diffusion Preliminaries} \label{SEC2}

\subsection{System Model and Problem Formulation}
We consider an uplink massive MIMO-OFDM system with $N_f$ subcarriers. The base station (BS) is equipped with $M_r$ elements and the user is equipped with $M_t$ elements. At the $k$-th subcarrier, the received signal at the BS is
\begin{equation}\label{eq.y}
    \mathbf{Y}_k = \mathbf{H}_k\mathbf{X}_k + \mathbf{Z}_k,\quad k=1,...,N_f,
\end{equation} 
where $\mathbf{X}_k\in {\mathbb C}^{M_t \times P}$ is the $P$-length transmitted pilot symbol. 
$\mathbf{z}_k \sim \mathcal{C} \mathcal{N}\left(0, \sigma_{z}^{2} \mathbf{I}_{M_{r}}\right)$ is the circularly symmetric complex Gaussian noise. Denote the $M_t$-point and $M_r$-point DFT matrices as $\mathbf{F}_{\rm t}\in\mathbb{C}^{M_t \times M_t}$ and $\mathbf{F}_{\rm r}\in\mathbb{C}^{M_r \times M_r}$, respectively. 
The angular domain representation of $\mathbf{H}_k$ is 
\begin{equation}
    \mathbf{H}_k = \mathbf{F}_{\rm r}\mathbf{H}^a_k\mathbf{F}_{\rm t}^{\rm T},
\end{equation}
in which $\mathbf{H}^a_k$ is the angular domain channel. Then the vectorization form of (\ref{eq.y}) is 
\begin{equation}\label{eq.yMh}
    \mathbf{y}_{k}=\mathbf{M}_k\mathbf{h}_k+\mathbf{z}_k, k = 1,\ldots, N_f,
\end{equation}
where $\mathbf{M}_k=(\mathbf{X}_k^{\rm T} \mathbf{F}_{\rm t} \otimes\mathbf{F}_{\rm r})\in{\mathbb C}^{M_r P \times M_r M_t}$, $\mathbf{y}_{k}={\rm vec}(\mathbf{Y}_{k})$, $\mathbf{h}_{k}={\rm vec}(\mathbf{H}^a_{k})$, and $\mathbf{z}_{k}={\rm vec}(\mathbf{Z}_{k})$. The channel matrices in different subcarriers originate from the same set of  physical propagation paths. Hence, the angular domain channel therefore exhibits a sparse structure with a common angular support. In the following derivation, we focus on a generic subcarrier and omit the subcarrier index for notational simplicity.

Based on the system model (\ref{eq.yMh}), the channel estimation problem can be formulated from a Bayesian perspective.
\begin{equation}
    \hat{\mathbf{h}} =  \underset{\mathbf{h}}{\operatorname{argmax}}\ p(\mathbf{h}|\mathbf{y}) = \underset{\mathbf{h}}{\operatorname{argmax}} \ p(\mathbf{y}|\mathbf{h})p(\mathbf{h}).
\end{equation}
In practice, the performance of the Maximum a posteriori (MAP) estimator critically depends on the accuracy of the prior distribution $p(\mathbf{h})$, especially in low signal-to-noise ratios (SNRs) or limited pilot scenarios where the likelihood term is insufficient for reliable channel recovery. In such cases, the prior plays a dominant role in regularizing the solution. Therefore, accurately characterizing the channel prior is essential for achieving reliable channel estimation.


\subsection{Conditional Diffusion Model Preliminaries}

Diffusion models construct a generative process by gradually perturbing data with Gaussian noise and learning to reverse this corruption process. 
Let $\mathbf{h}_0$ denote a clean data sample representing the channel realization. In conditional diffusion models, the generation process is guided by an auxiliary variable $\mathbf{c}$ that is statistically related to $\mathbf{h}_0$. In this work, the conditioning variable corresponds to the user location information, denoted as $\mathbf{c}=\mathrm{Loc}$, which provides prior information about the channel distribution. The pilot observation $\mathbf{y}$ will be incorporated later during the posterior inference stage to further refine the channel reconstruction.

Given a monotonic noise schedule $\beta_t\in (0,1)$ for $t=1,\ldots,T$ diffusion steps, the cumulative noise parameters are defined as $\alpha_t=1-\beta_t$ and $\bar{\alpha}_t=\prod_{i=1}^{t}\alpha_i$. The forward diffusion process generates a sequence of latent variables $\left\{\mathbf{h}_t\right\}_{t=1}^{\rm T}$ by progressively adding Gaussian noise according to
\begin{equation}
    q\left(\mathbf{h}_t \mid \mathbf{h}_{t-1}\right)=\mathcal{N}\left(\mathbf{h}_t ; \sqrt{\alpha_t} \mathbf{h}_{t-1},\left(1-\alpha_t\right) \mathbf{I}\right).
\end{equation}
As $t$ increases, the signal component gradually diminishes and $\mathbf{h}_T$ approaches an isotropic Gaussian distribution. Due to the Markov structure of the forward process, $\mathbf{h}_t$ can be expressed directly in terms of the original sample $\mathbf{h}_0$,
\begin{equation}
    \mathbf{h}_t=\sqrt{\bar{\alpha}_t} \mathbf{h}_0+\sqrt{1-\bar{\alpha}_t} \boldsymbol{\epsilon}_t,
\end{equation}
where $\boldsymbol{\epsilon}_t \sim \mathcal{N}(\mathbf{0}, \mathbf{I})$.

The reverse diffusion process aims to gradually remove the injected noise and recover the original data distribution conditioned on $\mathbf{c}$. The reverse transition is modeled as
\begin{equation}
    p_\theta\left(\mathbf{h}_{t-1} \mid \mathbf{h}_t,\mathbf{c}\right)=\mathcal{N}\left(\mathbf{h}_{t-1} ; \boldsymbol{\mu}_\theta\left(\mathbf{h}_t, t, \mathbf{c}\right), \sigma_t^2 \mathbf{I}\right),
\end{equation}
where $ \sigma_t^2=\frac{1-\bar{\alpha}_{t-1}}{1-\bar{\alpha}_t}\beta_t$. The mean $\boldsymbol{\mu}_\theta\left(\mathbf{h}_t, t, \mathbf{c}\right)$ can be derived as 
\begin{equation}
    \mu_\theta(\mathbf{h}_t, t, \mathbf{c})=\frac{1}{\sqrt{\alpha_t}}\left(\mathbf{h}_t-\frac{\beta_t}{\sqrt{1-\bar{\alpha}_t}}\boldsymbol{\epsilon}_{\theta}(\mathbf{h}_t, t, \mathbf{c})\right),
\end{equation}
where $\boldsymbol{\epsilon}_{\theta}$ is the trainable denoising function estimating the noise in the reverse process. Following the standard derivation of diffusion models, the training objective can be simplified to a noise prediction loss:
\begin{equation}
    \mathcal{L} = \underset{\mathbf{h}_0,t,\boldsymbol{\epsilon}_t}{\mathbb{E}} \|\boldsymbol{\epsilon}_{\theta}(\mathbf{h}_t, t, \mathbf{c})-\boldsymbol{\epsilon}_t \|_2^2.
\end{equation}

\section{Proposed Method}\label{SECsu}

\subsection{Environment-Aware Channel Estimation Formulation}

In most conventional approaches, the prior distribution $p(\mathbf{h})$ is assumed to be location-independent and characterized by global channel statistics. However, wireless channels are strongly coupled with the propagation environment, leading to significant spatial variability. 

\begin{figure}	
	\centering
	\begin{subfigure}[htbp]{3in}
		\centering
		\includegraphics[width=2in]{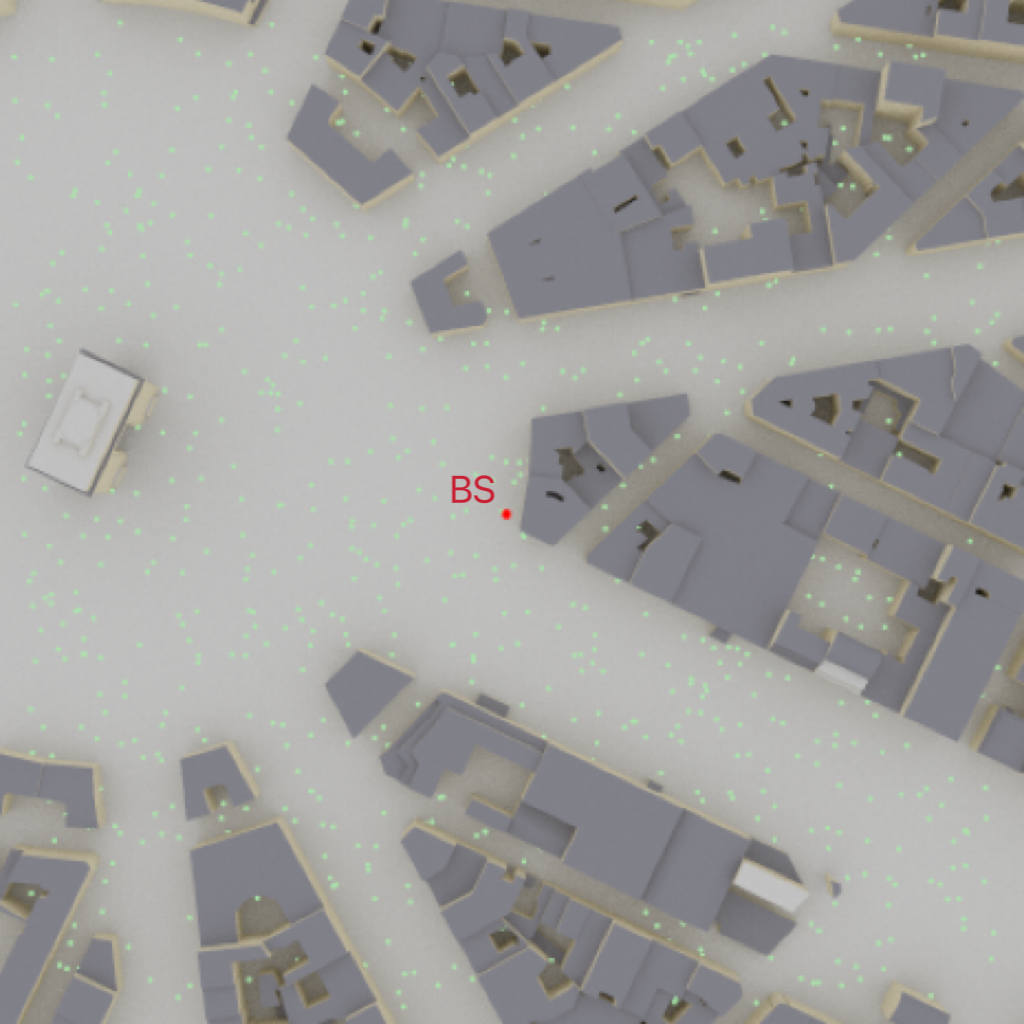}
		\caption{Simulation environment.}\label{fig.map}		
	\end{subfigure}
	\begin{subfigure}[htbp]{3in}
		\centering
		\includegraphics[width=3in]{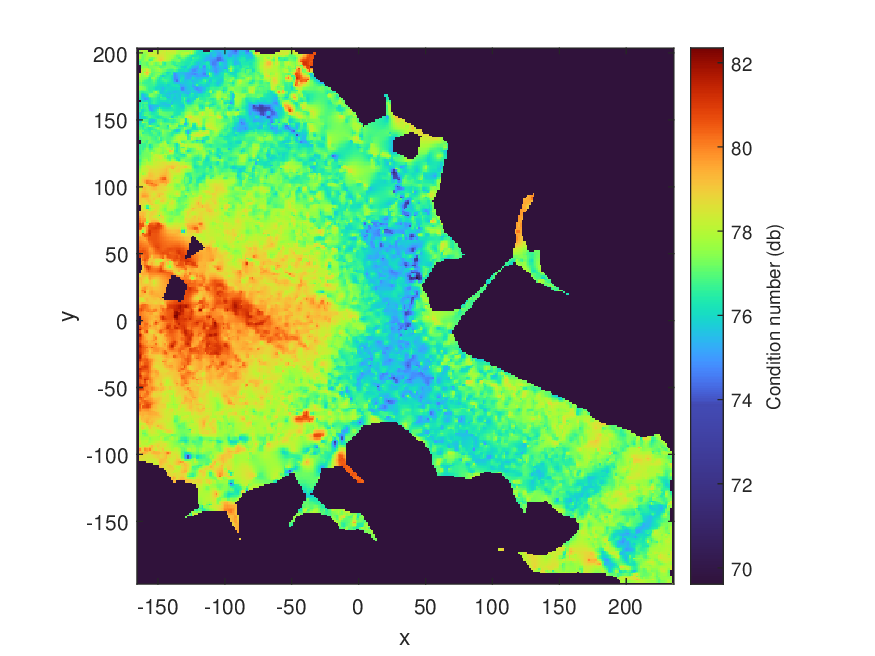}
		\caption{Spatial distribution of the channel condition number.}\label{fig.cond}
	\end{subfigure}
	\caption{Illustration of spatial variability of wireless channels.}\label{fig.1}
\end{figure}

To illustrate this, we consider a square area centered at the BS and generate ray-tracing wireless channels for users randomly distributed within this region as shown in Fig. \ref{fig.map}. The propagation environment is constructed based on a realistic map, and channel realizations are obtained for a large number of sampled user locations. The detailed ray-tracing simulation setup is provided in Section \ref{secSim}.
For each location, we compute the condition number of the corresponding channel matrix, which serves as a general indicator of channel characteristics\cite{yang2026complexity,yang2025achievable}. The spatial distribution of the channel condition number is visualized in Fig. \ref{fig.cond}. It can be observed that the channel condition number varies significantly across different spatial regions. 


Based on these observations, modeling the channel prior using a global location-independent distribution is insufficient. Instead, location information that reflects the spatial variability of channel statistics can provide a more accurate prior for channel estimation.

Considering the environment-induced spatial variability of wireless channels, the user location provides informative prior knowledge about the channel distribution. Let ${\rm Loc}$ denote the user location. In this paper, ${\rm Loc}$ is represented by a two-dimensional coordinate $(x,y)$, which specifies the horizontal position of the user in the considered area{\footnote{Although a two-dimensional coordinate $(x,y)$ is considered in this paper, the proposed framework can be readily extended to three-dimensional coordinates $(x,y,z)$. This only changes the dimensionality of the conditioning variable and does not affect the proposed algorithm or network architecture.}}. The environment-aware channel estimation can be formulated as 
\begin{equation}\label{eq.map_loc}
     \hat{\mathbf{h}} =  \underset{\mathbf{h}}{\operatorname{argmax}} \ p(\mathbf{h}|\mathbf{y},{\rm Loc}).
\end{equation}
However, accurately modeling the location-conditioned channel prior and performing inference on the high-dimensional posterior distribution remain challenging due to the complex propagation environment. In this work, we address this problem by leveraging diffusion models to learn the environment-aware channel distribution and perform channel estimation accordingly.

\subsection{Posterior Inference Algorithm}
Following the score-based posterior sampling framework, the channel estimation problem in (\ref{eq.map_loc}) can be solved through the following iterative posterior sampling procedure \cite{song2020denoising}
\begin{equation}\label{eq.ht-1}
    \mathbf{h}_{t-1} = \frac{1}{\sqrt{\alpha_t}}\left(\mathbf{h}_t+(1-\alpha_t)\nabla_{\mathbf{h}_t} {\log p_t(\mathbf{h}_t|\mathbf{y},{\rm Loc})} \right),
\end{equation}
where the gradient term $\nabla_{\mathbf{h}_t} {\log p_t(\mathbf{h}_t|\mathbf{y},{\rm Loc})}$ is the posterior score.
Since the received signal depends on the channel realization but not directly on the location, once $\mathbf{h}$ is given, the observation $\mathbf{y}$ is conditionally independent of ${\rm Loc}$ given $\mathbf{h}$. Therefore, according to Bayes' rule, the posterior  can be expressed as
\begin{equation}
    p(\mathbf{h}|\mathbf{y},{\rm Loc}) \propto p(\mathbf{y}|\mathbf{h}) p(\mathbf{h}|{\rm Loc}).
\end{equation}
Accordingly, the posterior score in (\ref{eq.ht-1}) can be represented as the sum of the conditional score and the likelihood score 
\begin{equation}
    \nabla_{\mathbf{h}_t} \log p_t(\mathbf{h}_t|\mathbf{y},{\rm Loc}) = \nabla_{\mathbf{h}_t} \log p_t(\mathbf{h}_t|{\rm Loc}) +\nabla_{\mathbf{h}_t} \log p_t(\mathbf{y}|\mathbf{h}_t).
\end{equation}
In practice, we adopt a modified posterior score as  \cite{dhariwal2021diffusion}:

\begin{equation}\label{eq.score_post}
\begin{aligned}
&\nabla_{\mathbf{h}_t} \log p_t(\mathbf{h}_t|\mathbf{y},{\rm Loc}) = \\
&\nabla_{\mathbf{h}_t} \log p_t(\mathbf{h}_t|{\rm Loc}) +w\nabla_{\mathbf{h}_t} \log p_t(\mathbf{y}|\mathbf{h}_t),
\end{aligned}
\end{equation}
where $w$ is a hyperparameter that balances the prior and likelihood terms, thereby improving denoising performance.

The likelihood score $\nabla_{\mathbf{h}_t} \log p_t(\mathbf{y}|\mathbf{h}_t)$ can be approximated following the derivation in \cite{zhou2025Generative}:
\begin{equation}\label{eq.score_likelihood}
\begin{aligned}
        &\nabla_{\mathbf{h}_t} \log p_t(\mathbf{y}|\mathbf{h}_t) = \\&\frac{1}{\sqrt{\bar{\alpha}_t}}\mathbf{M}^H\left(\frac{1-\bar{\alpha}_t}{\bar{\alpha}_t}\mathbf{M}\mathbf{M}^H+\sigma_z^2\mathbf{I} \right)^{-1}(\mathbf{y}-\frac{1}{\sqrt{\bar{\alpha}_t}}\mathbf{M}\mathbf{h}_t).
\end{aligned}
\end{equation}
The condition score can be approximately calculated using the denoising neural work \cite{song2020score}:
\begin{equation}\label{eq.score_prior}
    \nabla_{\mathbf{h}_t} \log p_t(\mathbf{h}_t|{\rm Loc}) \approx -\frac{1}{\sqrt{1-\bar{\alpha}_t}} \boldsymbol{\epsilon}_{\theta}(\mathbf{h}_t, t, \mathbf{c}).
\end{equation}


\begin{figure}[t]
\centering
\includegraphics[width=3.5in]{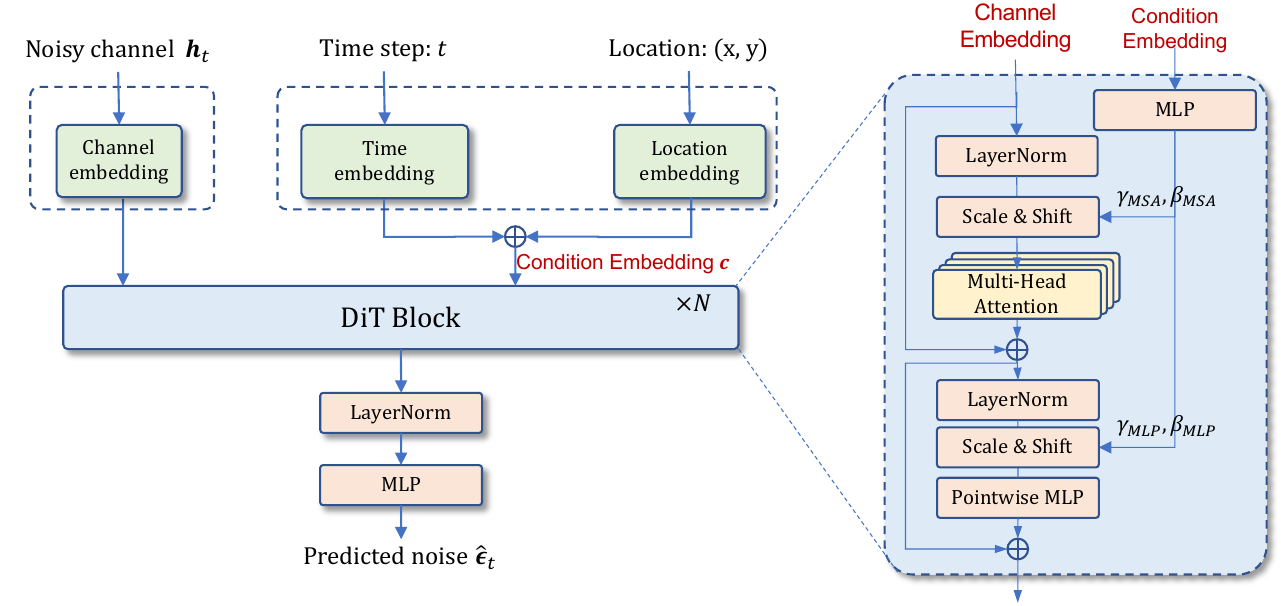}
\caption{Proposed Network Structure.}
\label{fig.network}
\end{figure}

\subsection{Network Structure}

We adopt a lightweight Transformer-based diffusion network inspired by the Diffusion Transformer (DiT) architecture. 
The network takes the noisy channel $\mathbf{h}_t$ as input and predicts the corresponding noise component $\boldsymbol{\epsilon}_{\theta}(\mathbf{h}_t, t, \rm{Loc})$. The model consists of an input embedding layer, stacked DiT blocks, and an output projection layer.

The complex channel $\mathbf{h}_t$ is first converted into real-valued features by separating the real and imaginary components and then projected to a hidden representation through a linear embedding layer. A sinusoidal positional encoding is added to preserve the structural information of the channel sequence.

The denoising process is conditioned on the diffusion timestep $t$ and the user location $\rm{Loc} = (x,y)$. Both variables are embedded into the same latent dimension using small multilayer perceptrons (MLP). The resulting embeddings are combined to form a unified conditioning vector that modulates the Transformer blocks.

The backbone network consists of $N$ stacked DiT blocks, each containing a multi-head self-attention module followed by a feed-forward network. Conditional information is injected through adaptive layer normalization, allowing the conditioning vector to modulate the intermediate feature statistics.

Finally, a linear projection maps the hidden representation back to the channel dimension to produce the predicted noise.

After obtaining the trained network, the proposed EA-DM channel estimation framework is given by Algorithm \ref{alg.ea-dm}.

\begin{algorithm}[b]
    \caption{EA-DM Channel Estimation Algorithm}
    \renewcommand{\algorithmicrequire}{\textbf{Input:}}
    \renewcommand{\algorithmicensure}{\textbf{Output:}}
    \begin{algorithmic}[1]
    
    \Require  $\mathbf{M}$, $\mathbf{y}$, $\rm Loc$, pre-trained network $\boldsymbol{\epsilon}_{\theta}$, noise schedule $\{\beta_t\}_{t=1}^{T}$
    \Ensure  $\hat{\mathbf{h}}$.
    \State Initialize $\mathbf{h}_{T}\sim\mathcal{CN}(0,\mathbf{I})$ and calculate $\alpha_t=1-\beta_t$, $\bar{\alpha}_t=\prod_{i}^{t}\alpha_i, \forall {t}$.
    \For{$t = T:-1:1$}
    \State Calculate condition score $\nabla_{\mathbf{h}_t} \log p_t(\mathbf{h}_t|{\rm Loc})$ by (\ref{eq.score_prior}).
    \State Calculate likelihood score $\nabla_{\mathbf{h}_t} \log p_t(\mathbf{y}|\mathbf{h}_t)$ by (\ref{eq.score_likelihood}).
    \State Calculate the posterior score using (\ref{eq.score_post}).
    \State Update $\mathbf{h}_{t-1}$ by (\ref{eq.ht-1}).
    \State $\hat{\mathbf{h}}=\mathbf{h}_0$
    \EndFor
    \end{algorithmic}\label{alg.ea-dm}
\end{algorithm}
\section{Numerical Results}\label{secSim}

In the simulations, the channel dataset is generated using the ray-tracing module of the open-source wireless communication simulator Sionna \cite{sionna}. Specifically, as shown in Fig. \ref{fig.map}, the pre-defined urban environment of Munich is adopted as the propagation scene, which provides detailed geometric information of buildings and streets. 
Specifically, user locations are randomly generated within a $200{\rm m} \times 200{\rm m}$ region centered at the BS. The height of each user terminal is fixed at $1.5\rm{m}$. The BS employs an $8\times8$ uniform plane array (UPA) and each UE employs a $4\times4$ UPA, i.e., $M_r=64, M_t=16$. The carrier frequency is $28\rm{GHz}$. The OFDM system uses $32$ subcarriers with a subcarrier spacing of $15\rm{kHz}$. Based on the above configuration, the final dataset consists of channel realizations generated for $N_{train}=13{,}800$ randomly sampled user locations.


We evaluate channel estimation performance through the normalized mean squared error (NMSE), which is defined as ${\rm NMSE}=\frac{\mathbb{E}[\|\hat{\mathbf{h}}-\mathbf{h}\|_2^2]}{\mathbb{E}[\|\mathbf{h}\|_2^2]}$. 
Three benchmark schemes are considered, i.e., regularized least squares (LS), linear minimum mean square error (LMMSE), and the CNN-DM in \cite{zhou2025Generative}.
\begin{itemize}
    \item \textbf{Regularized LS:} As we consider the underdetermined case, i.e., $P<M_t$, we adopt the regularized LS to obtain $\hat{\mathbf{h}}=(\mathbf{M}^{\rm H}\mathbf{M}+\sigma_z^2\mathbf{I})^{-1}\mathbf{M}^{\rm H}\mathbf{y}$.
    \item \textbf{LMMSE:} The sample covariance is computed via all training data, $\mathbf{R}_h=\frac{1}{N_{train}}\sum_{i=1}^{N_{train}}\mathbf{h}\mathbf{h}^H$. Then the estimated channel is $\hat{\mathbf{h}}=\mathbf{R}_h\mathbf{M}^{\rm H}(\mathbf{M}\mathbf{R}_h \mathbf{M}^{\rm H}+\sigma_z^2\mathbf{I})^{-1}\mathbf{y}$.
    \item \textbf{CNN-DM in \cite{zhou2025Generative}:} The diffusion model based on CNN performs MAP channel estimation, without considering environmental priors.
\end{itemize}

For both EA-DM and the CNN-DM, the number of diffusion steps is set to $T=100$. For network training, both models are trained for $500$ epochs with a batch size of $128$ using the Adam optimizer with a learning rate of $10^{-3}$.

\begin{figure}[tp]
\centering
\includegraphics[width=3.1in]{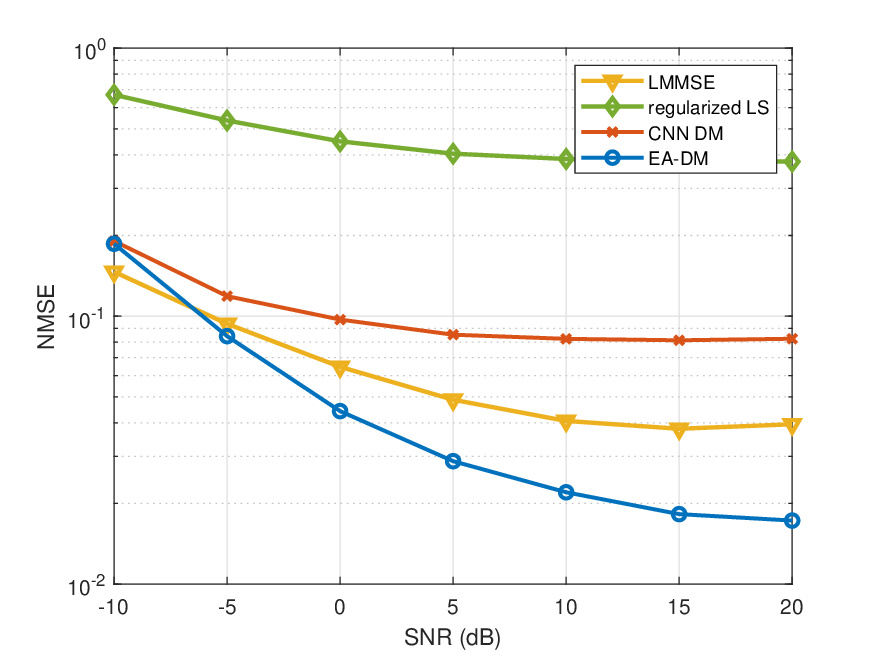}
\caption{NMSE versus SNR for different channel estimation schemes.}
\label{fig.nmse_snr}
\end{figure}

\begin{figure}[tp]
\centering
\includegraphics[width=3.1in]{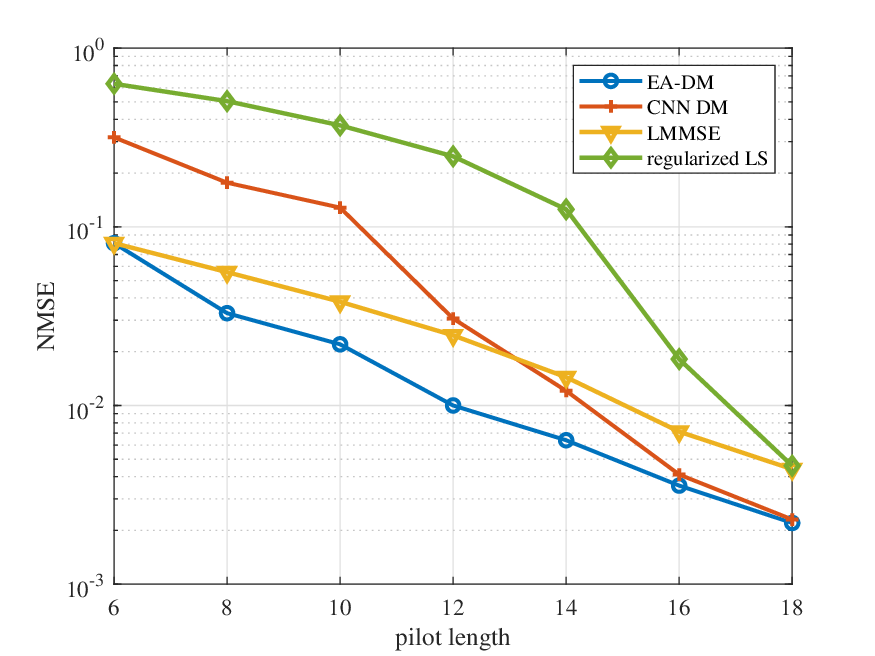}
\caption{NMSE versus pilot length for different channel estimation schemes with ${\rm SNR}=20{\rm dB}$.}
\label{fig.nmse_p}
\end{figure}


Fig. \ref{fig.nmse_snr} illustrates the NMSE performance versus SNR for different channel estimation schemes. The pilot length is set as $P=10$. The regularized LS  consistently underperforms due to the insufficient pilot dimension.  LMMSE provides a strong linear baseline by exploiting global second-order channel statistics, but its Gaussian covariance-based prior cannot fully characterize the channel distribution. CNN-DM learns a data-driven but location-independent prior  $p(\mathbf{h})$, which effectively represents an averaged channel distribution across different spatial regions. However, in this scenario, the CNN-DM is less effective than the stable covariance-based shrinkage of LMMSE.
In contrast, the proposed EA-DM leverages environment-aware information to capture the spatial structure of the channel distribution, enabling a more accurate prior and leading to improved estimation accuracy.

Fig. \ref{fig.nmse_p} shows the NMSE performance versus pilot length, with the SNR fixed at $20$dB. The performance of all methods improves with increasing pilot length. When the pilot length is short, the observation provides limited likelihood information and the estimation relies heavily on prior regularization; in this regime, CNN-DM is limited by its global location-independent prior, while LMMSE remains relatively robust due to covariance-based shrinkage.
As the pilot length increases, the observation becomes more informative and the reliance on prior information decreases. Consequently, the performance gap between EA-DM and CNN-DM becomes smaller, since both methods can benefit from the increasingly accurate observations. Nevertheless, EA-DM maintains the best overall performance, demonstrating the effectiveness of incorporating environment-aware information for channel estimation.

\section{Conclusions} \label{SEC6}
In this paper, we proposed the EA-DM for channel estimation in massive MIMO-OFDM systems. By exploiting the spatial variability of wireless channels, a conditional diffusion model was trained to learn the relationship between user location and channel state information, providing a location-conditioned channel prior for channel estimation. Based on the learned prior, a posterior inference algorithm was developed to incorporate pilot observations into the reverse diffusion process, enabling Bayesian channel estimation that jointly leverages location information and measurement data. Simulation results based on ray-tracing channel datasets demonstrate that the proposed method achieves improved estimation accuracy compared with conventional and existing learning-based approaches.

\bibliographystyle{ieeetr}
\bibliography{all}

\end{document}